\documentclass[twocolumn,amssymb,floatfix,aps,prb,showpacs,citeautoscript]{revtex4}

\usepackage{epsfig} 
\usepackage{amsmath}
\usepackage{mleftright}\mleftright
\usepackage{hyperref}
\usepackage{times}

\usepackage{graphicx}

\begin{document} 

\title{Spatial BCS-BEC crossover in  superconducting p-n junctions}

\author{A. Niroula} 

\address{Jacobs University, 
  Campus Ring 1, 28759 Bremen, Germany}

\author{G. Rai} 
\address{Department of Physics and Astronomy, University of Southern California, Los Angeles, CA 90089-0484, USA}

\author{ S. Haas} 
\address{Department of Physics and Astronomy, University of Southern California, Los Angeles, CA 90089-0484, USA}

\address{Jacobs University, 
  Campus Ring 1, 28759 Bremen, Germany}

\author{S. Kettemann} 

\address{Jacobs University, 
  Campus Ring 1, 28759 Bremen, Germany}
\address{ Division of Advanced
  Materials Science POSTECH, San 31, Hyoja-dong, Nam-gu, Pohang 790-784, South Korea}

%
%
%

\date{\today}

\begin{abstract} 
We present a  theory  of     
   superconducting p-n junctions. To this end, we  consider a two band model of  doped bulk 
 semiconductors 
 with attractive interactions between the charge carriers 
 and derive  the superconducting  order parameter, the quasiparticle  density of states and the 
 chemical potential as a function of the semiconductor
  gap $\Delta_0$ and  the doping level $\varepsilon$. 
  We verify  previous results for the quantum phase diagram for a system with constant density of states in the conduction and valence band, which show    BCS-Superconductor to Bose-Einstein-Condensation (BEC) and BEC to Insulator  transitions  as  function of  doping level and the size of the band gap. Then, we extend this formalism to a density of states which is more realistic for 3D systems
   and derive the corresponding quantum phase diagram, where we find that a BEC phase can only exist for 
    small band gaps $\Delta_0 < \Delta_0^*$.  For larger band gaps, we find rather a direct transition from an 
     insulator to a BCS phase. 
   Next, we apply this theory  to  study the properties of    
   superconducting p-n junctions. We derive 
 the spatial variation of the superconducting
order parameter  along the p-n junction. As the potential difference across the junction leads to energy band bending, 
 we find a   spatial  crossover 
 between a BCS and BEC condensate, as the density of charge carriers changes across the p-n junction. For the 2D system, we find two possible regimes, when the bulk is in a BCS phase, a BCS-BEC-BCS junction with a single BEC layer in the space charge region, and a BCS-BEC-I-BEC-BCS junction with two layers of BEC condensates separated by an insulating layer. In 3D we find that there can also be a conventional BCS-I-BCS junction for semiconductors with band gaps exceeding $\Delta_0^*$. 
   Thus, we find that there can be BEC layers  in the well controlled setting of doped semiconductors, where the doping level can be varied to  change and control the thickness of BEC and insulator layers,
    making Bose Einstein Condensates  thereby possibly accessible to experimental transport and optical studies
     in solid state materials. 
\end{abstract}


\maketitle


\section{Introduction} 
  The existence of a superconducting state below a critical temperature $T_c$
  is not restricted to materials which are typical metals at higher temperatures,
  but can also occur in materials 
   that are known to be  semiconductors.\cite{Cohen,Hanke} For example, superconductivity has  
     been observed at   doping concentrations  as small as $4 \times 10^{17} cm^{-3}$  in SrTiO$_{3}$,  with a critical temperature of $T_c =0.1\thinspace K$ \cite{lin}, and  in a wide range of doped semiconductors, such as B-doped diamond
   \cite{Ekimov2004,Blase2004,Bustarret2004}
    and  in doped silicon under high pressure, \cite{Bustarret2006} with critical temperatures up to $T_c =10\thinspace K$.

   The BCS theory of superconductivity\cite{BCS1,BCS2,BCS3}, 
   can be extended and applied to  such materials.  Eagles \cite{Eagles}  has solved the BCS equations
  within  a single-band semiconductor model and found a crossover to a BEC condensate as the doping concentration is lowered.  There,  the charge carriers form local pairs which condense  into a Bose-Einstein condensate at low temperatures\cite{Leggett1980}. Nozieres and Pistolesi \cite{Nozieres} have  extended this theory to a two-band semiconductor model and studied the superconducting-insulator transition as a function of the semiconductor energy gap,  for   a constant density of states in each band, as well as for a particular non constant density of states with an exponential dependence on energy.  BCS-BEC crossover in multiband systems has been 
  further studied in Refs. 
  \onlinecite{Chubukov2016}, \onlinecite{Loh2016} and
  \onlinecite{Yerin2019}.
   Experimemtally,  the BCS-BEC crossover has
   first been studied in artificial atom systems 
\cite{Regal2004},
\cite{Zwierlein2004}.
   Recently,  the BCS-BEC  has been experimentally  studied 
  in the  Fe-Based superconductor 
      $Fe_{1+y} Se_x Te_{1-x}$ \cite{Rinott2017} by chemical  variation of the doping level  and    in single-crystalline lithium-intercalated layered nitrides by gate controlled doping \cite{Nakagawa2018}.  Superconductivity has been discovered in magic angle 
      twisted bilayer graphene at low carrier concentrations, which is  tunable by gate controled doping\cite{Cao2018}  and might open another venue to study the BCS-BEC crossover experimentally.
  
  Junctions between p- and n-doped semiconductors form the basic element of semiconductor devices whose  rectifying behavior is based on the energy band bending and on the different  majority 
 charge carriers, holes and electrons, respectively on either side of the junction.  
   As superconductivity has been  observed  both in p- and n-doped semiconductors, intriguing questions arise about the physical properties of superconducting p-n junctions\cite{Mannhart}: how does the superconducting order parameter vary spatially across the junction? Does a p-n junction form a Josephson contact, and how large is  the supercurrent across the p-n junction?  
    Such questions have been explored for 
   $YBa_2Cu_3O_7/Nd_{1-x}Ce_xCu_2O_4$  junctions\cite{Takeuchi1995},  with an estimated depletion width of  less than $1 \thinspace nm$ \cite{Mannhart}, for the  p-type superconductor YBa2Cu3O (YBCO) over the n-type superconducting cuprate Pr2CexCuO4 (PCCO)\cite{Wu2007}, as well as   for iron pnictide p-n junctions, where the redistribution of charges could possibly lead to the suppression of the local superconducting order parameter near the interface for both single crystals. This may play a role in the junction formation itself  \cite{Zhang2009}.
   The superconductivity  in magic angle 
      twisted bilayer graphene  has been obtained both for electronic and hole gate controlled  doping\cite{Cao2018},  which might allow to form 
      superconducting p-n junctions from twisted bilayer graphene.
  
  Here, we  study  superconducting p-n junctions within a two-band model, based on a self-consistent solution of the BCS equations, the Poisson equation and the particle number conservation. In the next section, we first review the two-band theory of superconductivity  for a constant density of states. Then, we  generalize it to a more realistic
    three-dimensional  density of states.  We derive the pairing amplitude, the chemical potential, 
   the quasiparticle density of states and the coherence length $\xi$ as functions of the semiconductor band gap $\Delta_0$ and the doping level $\varepsilon$. We identify the crossover between superconductivity (SC) and Bose-Einstein condensation (BEC) and derive the 
    corresponding phase diagram in the $\varepsilon$-$\Delta_0$ parameter space. 
    Based on this model, in section III we derive the properties of a superconducting p-n-junction homojunction (with same parent material on both sides of the junction), in particular the spatial dependence of the order parameter, the quasiparticle excitation energy and the 
     pairing coherence length across the p-n junction. 
    
  \section{Two-band theory of superconductivity}
\begin{figure}[h]
\includegraphics[width=4.5cm]{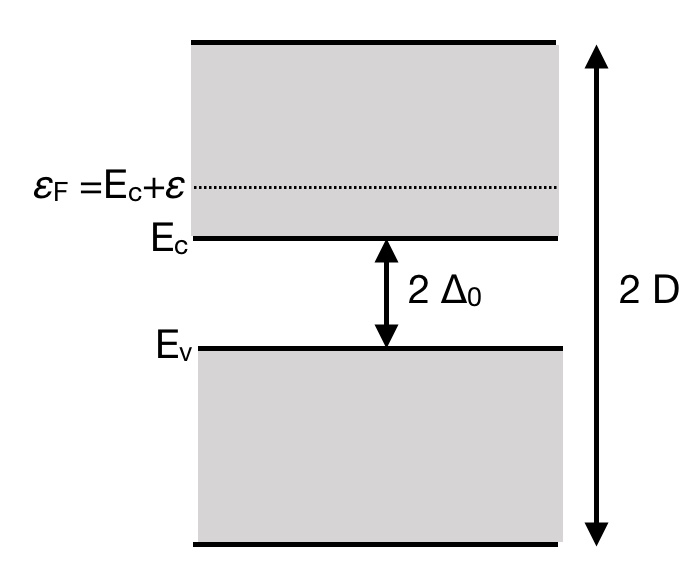}
   \includegraphics[width=4.cm,angle=0]{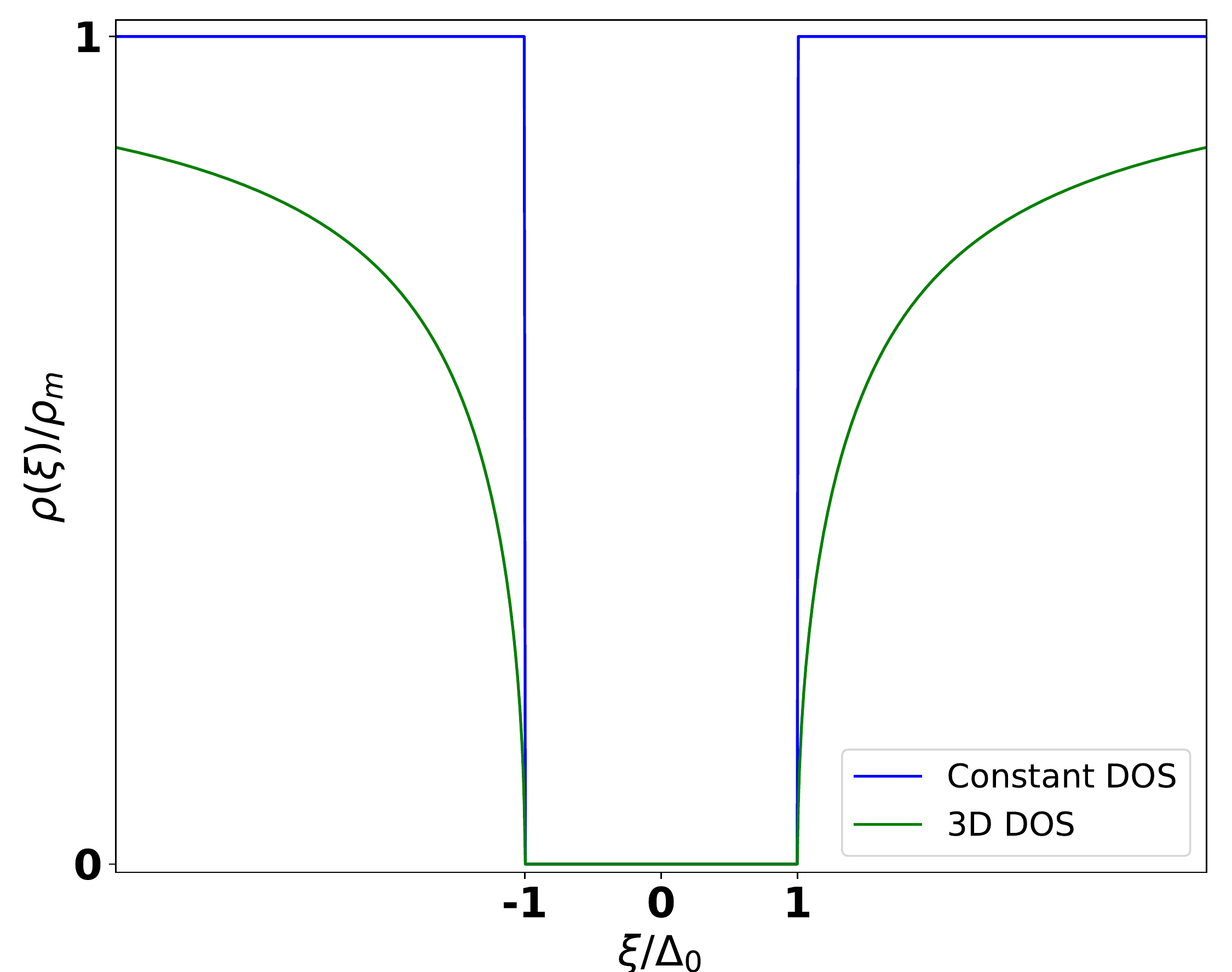}
\caption[Energy Scale and DOS]{Left: two-band model
with valence band edge $E_V= - \Delta_0$, and conduction band
 edge $E_C = \Delta_0$,  semiconductor band gap $ 2 \Delta_0$, total band width $2D$. Energy range with attractive pairing  $2\omega_D$ around Fermi energy $\varepsilon_ F = E_c + \varepsilon$ with doping level $\varepsilon$.  Right: model densities of states  as a function of energy: 2D  DOS (blue) and 3D DOS (green).}
 \label{fig:1} 
\end{figure}

 In order to derive the superconducting order parameter 
  $\Delta$ and the chemical potential $\mu$, we need to solve the BCS self-consistency equation along
  with  the equation for the 
  conservation of particle number $N$. 
The particle number conservation at $T=0$ gives\cite{Nozieres}:
\begin{eqnarray}\label{ParticleNumberConservation}
2 \int d \xi_k \rho\left(\xi_k\right) v_k^2
 =N= 2\int^{\varepsilon_{\rm F}} d \xi_k \rho\left(\xi_k\right),
\end{eqnarray}
 where  $\rho\left(\xi_k\right)$ is the density of states,
 $v_k^2 = \left(1-(\xi_k-\mu)/E\left(\xi_k\right)\right)/2$
 with electron energy dispersion $\xi_k$,
$E(\xi_k) = \sqrt{(\xi_k-\mu)^2 + \Delta^2}$ is the quasiparticle energy.
$ \varepsilon_{\rm F}$ is the Fermi energy at $T=0K$.
 At $T = 0$, there are no thermally excited charge carriers.
  Doping introduces additional electrons or holes. However, in the dilute doping limit, 
  electrons and holes are trapped at low temperature by the donor and acceptor atoms, respectively.
  As the concentration of donor atoms $N_D$ or acceptor atoms $N_A$  increases, their eigenstates hybridize and eventually delocalize into impurity bands, which at larger doping concentrations merge with the conduction or valence band, respectively. 
   Here, we model the doping in a simplified way by
   a continuous variation of the Fermi energy, for  donor doping by  $ \varepsilon_F = E_C + \varepsilon_n$ 
   and for  acceptor  doping by  $ \varepsilon_F = E_V - \varepsilon_p$
    (see Fig.  \ref{fig:1}).
    
   These doping parameters are   related to 
the donor concentration $N_D$ and  the acceptor concentration $N_A$, respectively,
 for the 2D DOS via $N_{D}=2 \rho_{0} \varepsilon_{n}, N_A= 2 \rho_{0} \varepsilon_{p}$,
 where the factor 2  accounts for  the spin degeneracy. 
 For the 3D DOS, one finds 
$N_{D}=2 \frac{2}{3}\rho_{0}\thinspace\varepsilon_{n}^{\frac{3}{2}}, N_{A}=2 \frac{2}{3}\rho_{0}\thinspace\varepsilon_{p}^{\frac{3}{2}}.$

 The BCS weak coupling theory gives for $T=0K$ the self-consistency equation  for the 
  order parameter $\Delta$,
\begin{equation}\label{SelfConsistency}
1=\frac{U}{2}\int_{\mu-\omega_{D}}^{\mu+\omega_{D}}d\xi_k\frac{\rho\left(\xi_k\right)}{\sqrt{\left(\xi_k-\mu\right)^{2}+\Delta^{2}}},
\end{equation}
 where $U$ is the attractive interaction strength, and $ 2 \omega_{D}$ is the size of the typical energy window around the chemical potential $\mu$ where the effective interaction is attractive.\\
 
    {\it Quasiparticle density of states.~}
  The quasiparticle density of states is 
   defined by 
    \begin{eqnarray}
    N(E) = - \frac{1}{\pi} Tr Im \hat{G}_E,
    \end{eqnarray}
    where $E$ is the quasiparticle excitation energy relative to the 
     chemical potential $\mu$, and $\hat{G}_E$ is the quasiparticle propagator.    
     
\begin{figure}[h] 
\includegraphics[width=5cm,angle=0]
{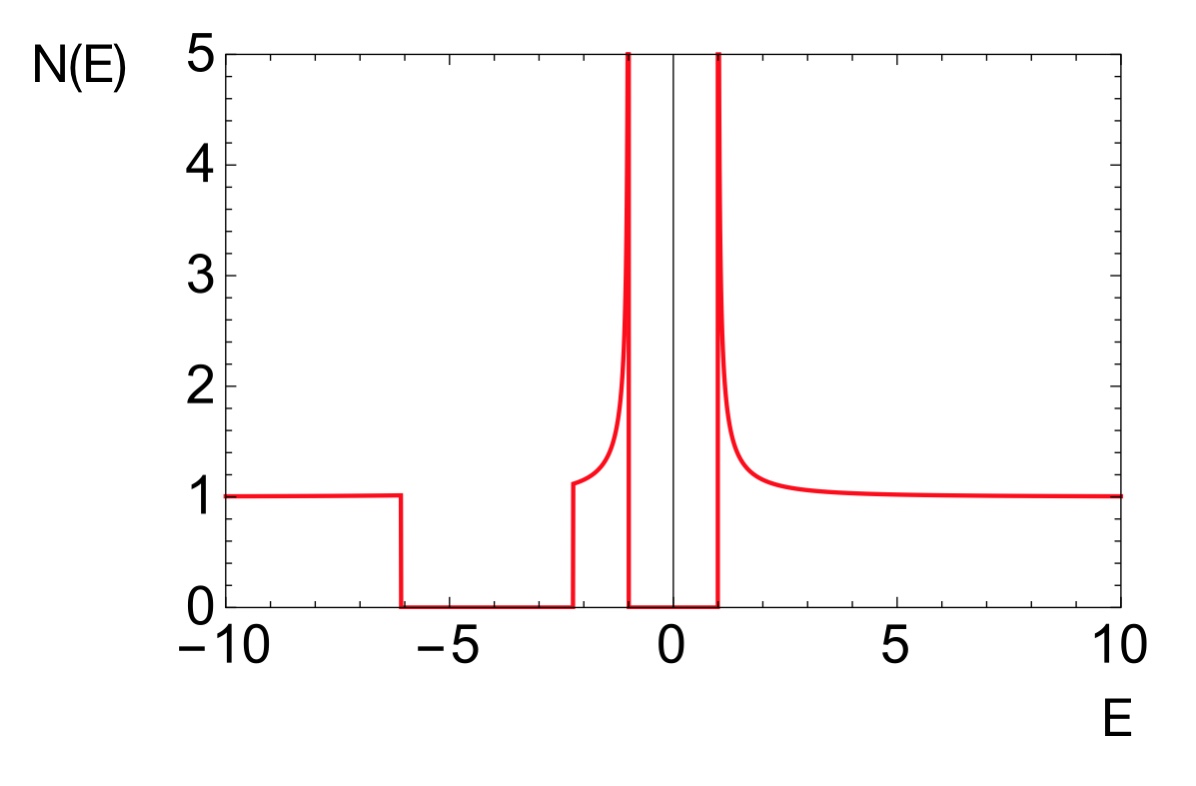}
\includegraphics[width=5cm,angle=0]
{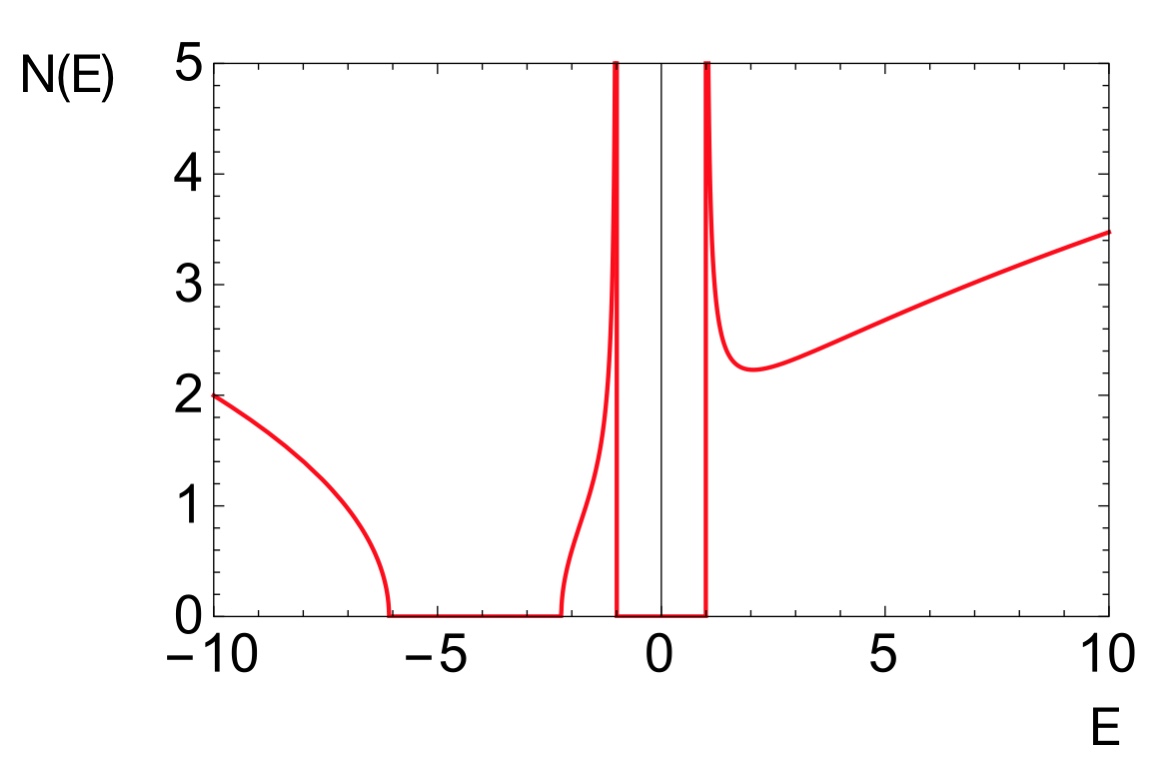}
\includegraphics[width=5cm,angle=0]{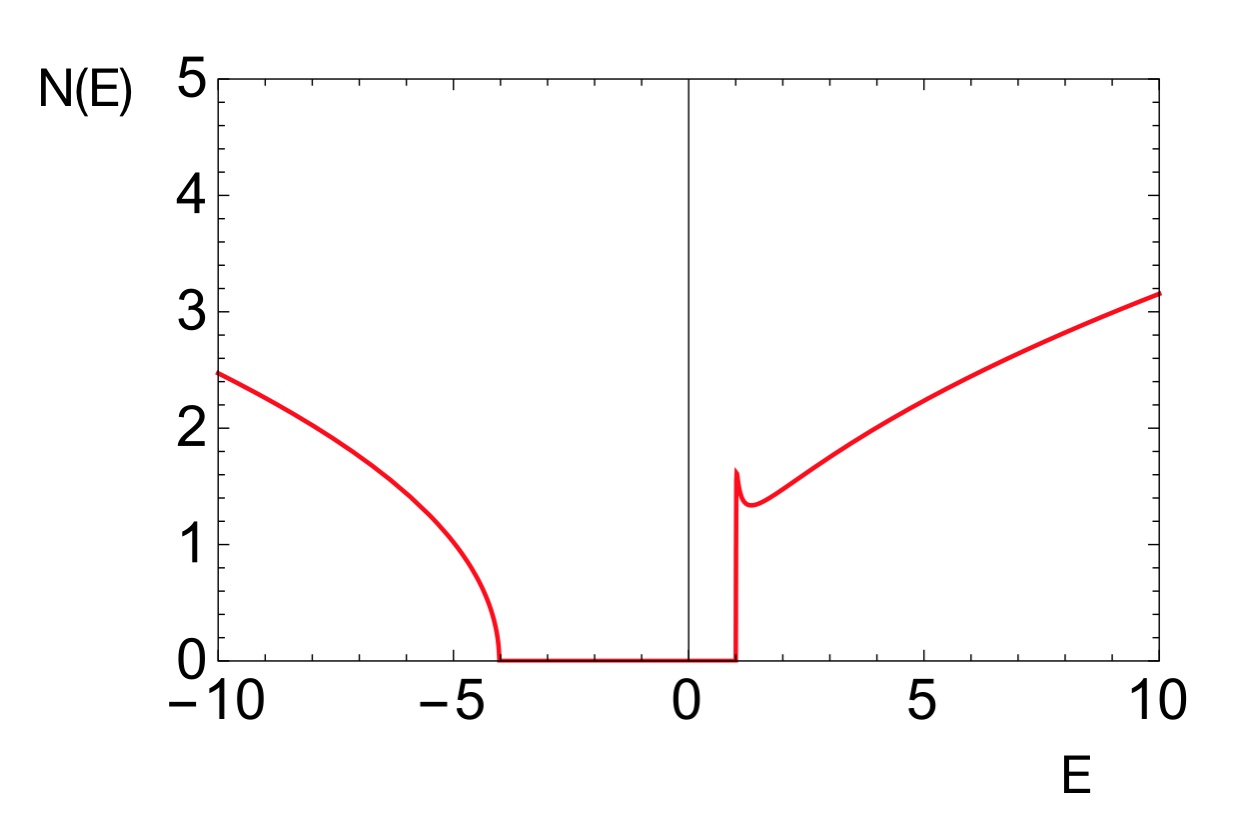}
\caption{2D (Top) and 3D (Center) quasiparticle density of states, Eq. (\ref{sdos}), as a function of quasiparticle energy $E$, when the chemical potential is in the conduction band. (Bottom) quasiparticle density of states as function of quasiparticle energy E for the 3D DOS, when the chemical potential is in the semiconductor band gap.
} \label{fig:2}
\label{fig:2}
\end{figure}

     Noting that in the presence of the pairing gap $\Delta$, the propagator is given by\cite{schrieffer} 
       $G_E (\xi_k) = (E+i \delta +\xi_k-\mu))/((E+i\delta)^2 - \Delta^2 - (\xi_k-\mu)^2),$
       and thus we get via complex integration
       \begin{eqnarray} \label{sdos}
         N(E) = Re [ \rho (\mu + \sqrt{E^2-\Delta^2}) \frac{|E|}{\sqrt{E^2-\Delta^2}} ].
       \end{eqnarray}
       
       When the chemical potential is within a band, e.g. in the conduction band, 
       $\Delta_0 < \mu < D$, the quasiparticle density of states $N(E)$ diverges at $ E=\pm \Delta$,  the coherence peak, 
        and is zero for smaller energies, so that $\Delta$ is the quasiparticle gap (see Fig. \ref{fig:2}(top) and (center)). 
         Remarkably, in the case when the chemical potential is in  the semiconductor gap, 
          $ -\Delta_0 < \mu < \Delta_0$, the quasiparticle density of states $N(E)$ does not diverge for any E, see Fig. \ref{fig:2}(bottom) , but it is still peaked.
          This is an indication  that the  system is in a  Bose-Einstein  condensate, as we discuss below. 
          Moreover, the quasiparticle gap 
          is then  enhanced to $\Tilde{\Delta} = \sqrt{\Delta^2 + (\Delta_0-|\mu|)^2} > \Delta$ exceeding the pairing order parameter $\Delta$. \\

{\it BCS-BEC Crossover.~}
There is a crossover from BCS superconductivity to Bose-Einstein condensation (BEC)  as the concentration of charges carriers is lowered  by decreasing the doping level $\varepsilon$\cite{Nozieres}.  Let us study  this 
  BCS-BEC crossover in more detail. One way to distinguish between BCS and BEC is to measure the coherence length $\xi$ of the condensate pairs.  
  When $ \xi > \lambda_F$, where $\lambda_F$ is the Fermi wave length,  many electron pairs overlap with each other, which is typical for a superconducting condensate.  When $ \xi < \lambda_F,$ however, the electron pairs do not overlap, but they instead form well-defined bosons which condense  below the transition temperature $T_c$.
   Therefore, let us next calculate $\xi$ in the two-band model. 
   $\xi$ can be derived by calculating the expectation value of the distance between
    two electrons  with opposite spin in the  ground state
    \begin{equation}
        \xi^2 = \int dr r^2 g(r)/ \int dr g(r).
    \end{equation}
    Here, $g(r)$ is the pair correlation function in the ground state, defined by 
    \begin{equation}
        g(r)=|\langle \psi | \psi_+^{\dagger}(r) \psi_-^{\dagger}(0) |\psi\rangle|^2,
    \end{equation}
    where $|\psi\rangle$ is the  BCS trial ground state given by 
       \begin{equation}
        |\psi\rangle =\prod_{k} \left(u_k + v_k c_{k+}^{\dagger} c_{k-}^{\dagger} \right) |0\rangle.
    \end{equation}
    Here $c_{k\alpha}^+$ are the fermion creation operators in a state with momentum $k$ and spin $\alpha = \pm$. $|0\rangle $ is the vacuum state, and
    $2 u_k v_k =\Delta/\sqrt{(\xi_k-\mu)^2 +\Delta^2}$.
    The electron field operators are given by $\psi_{\alpha}^+(r) = \sum_k e^{\i k r} c_{k \alpha}^+.$
    Thereby we find 
    \begin{equation} \label{xi}
        \xi^2 =- \sum_k u_k v_k \nabla_k^2 u_k v_k/ \sum_k u_k^2 v_k^2.
    \end{equation}
We will calculate $\xi$ below for the two-band model explicitly. 

        \subsection{Two-band model with 2D DOS}
         Let us first review the theory for the two-band model with a
         constant density of states $\rho$ in both the valence and the conduction band,
    separated by an energy gap $2 \Delta_0$ as 
shown in Fig.\ref{fig:1}.     
        This corresponds to  a two-dimensional 
         system, as considered in Ref. \onlinecite{Nozieres}.
        As we will mostly be interested to understand 
         the BCS-BEC crossover limit where the Fermi energy relative to the band edge is small, we will first  follow Ref. \onlinecite{Nozieres} in assuming that $\omega_D$ is a  large energy scale. 
         This means that we assume that the  electron-electron  interaction is attractive in both bands,  so that  we can set $\omega_D = D$.
         Thereby, 
the BCS self-consistency equation  simplifies to 
\begin{align}\label{DopedSemiConductorSelfConsistency1}
1=\frac{\rho U}{2}\left( \int_{\Delta_{0}}^{\omega_{D}}  + \int_{-\omega_{D}}^{-\Delta_{0}} \right) d\xi_k\frac{1}{\sqrt{\left(\xi_k-\mu\right)^{2}+\Delta^{2}}},
\end{align}
which  gives
\begin{align}\label{DopedSemiConductorSelfConsistencySimplified11}
\frac{2}{\rho U}=\ln\left(\frac{\omega_{D}-\mu+\sqrt{\left(\omega_{D}-\mu\right)^{2}+\Delta^{2}}}{\Delta_{0}-\mu+\sqrt{\left(\Delta_{0}-\mu\right)^{2}+\Delta^{2}}} \notag \right.
\\\left.
\cdot\frac{-\Delta_{0}-\mu+\sqrt{\left(-\Delta_{0}-\mu\right)^{2}+\Delta^{2}}}{-\omega_{D}-\mu+\sqrt{\left(-\omega_{D}-\mu\right)^{2}+\Delta^{2}}}\right).
\end{align}

For the limiting case of a gapless, metallic system, i.e., $\Delta_0=0,$  
we can express the interaction factor $\rho U$
 in terms of the  superconducting order parameter
 $\Delta_{m}$ via 
 \begin{equation}\label{ConstDOSSelfConsistencySimplified}
\Delta_{m}\equiv \Delta(\Delta_0=0)=2\omega_{D}\exp\left(\frac{-1}{\rho U}\right),
\end{equation} 
 and rewrite Eq. (\ref{DopedSemiConductorSelfConsistencySimplified11}) as
\begin{align}\label{ConstDOSMainEqn1}
\Delta_{m}^{2}=\left[\left(\Delta_{0}-\mu\right)+\sqrt{\left(\Delta_{0}-\mu\right)^{2}+\Delta^{2}}\right] \notag
\\
\cdot \left[\left(\Delta_{0}+\mu\right)+\sqrt{\left(\Delta_{0}+\mu\right)^{2}+\Delta^{2}}\right].
\end{align}
Particle conservation  in the  doped semiconductor
implies that the number of particles does not change as superconductivity sets in. 
Therefore, we  need to ensure the equality 
 between the number of particles in the normal and in the superconducting state:
 For an n-doped semiconductor, electrons are released into the conduction band by donor atoms.
  We model this  by adding an extra number of electrons $\delta N$, 
 which for a constant density of states in the conduction band can be written as
  $\delta N = 2 \rho \varepsilon$. Here, $\varepsilon = \varepsilon_{\rm F}- \Delta_0$ is the Fermi energy measured from the conduction band edge $\Delta_0.$
\begin{align}\label{ParticleConservationDopedSC}
\underset{\delta N}{\underbrace{2\rho \varepsilon}}+2\rho\int_{-D}^{-\Delta_{0}}d\xi_k=
\left( \int_{-D}^{-\Delta_{0}} +\int_{\Delta_{0}}^{D}
\right) d \xi_k
\left(1-\frac{\xi_k-\mu}{E\left(\xi_k\right)}\right).
\end{align}
 Here,   $2 D$ represents the total  bandwidth of the semiconductor. For
large $D+\mu\gg\Delta$ and $D-\mu\gg\Delta$, integration gives
\begin{align}\label{ParticleConservationDopedSCSimplified1}
2\varepsilon=2\mu-\sqrt{\left(\Delta_{0}+\mu\right)^{2}+\Delta^{2}}+\sqrt{\left(\Delta_{0}-\mu\right)^{2}+\Delta^{2}}.
\end{align}
Eqs. (\ref{ConstDOSMainEqn1}) and  (\ref{ParticleConservationDopedSCSimplified1}) are the set of  equations that describe the BCS superconducting  state of 
semiconductors with constant density of states.
   
\begin{figure}[t!] 
\includegraphics[width=7cm]{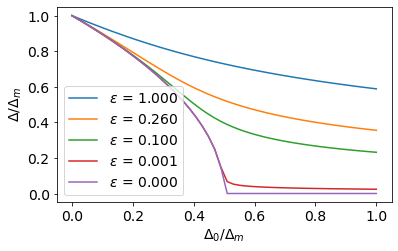}
\includegraphics[width=7cm]{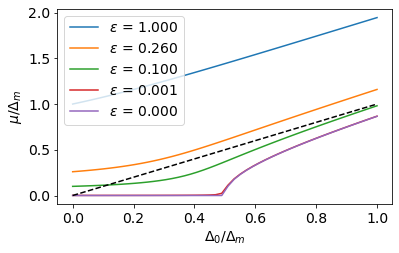}
\caption{ 
Top: superconducting order parameter $\Delta$, Bottom: chemical potential $\mu$ as functions of the semiconductor band gap $\Delta_{0}$, at different doping levels $\varepsilon$
 for $T=0\thinspace K$ and  2D constant density of states.  $\Delta_m$ is the superconducting order parameter for the metallic ($\Delta_0=0$) case,  Eq. (5). The dashed line is the conduction band edge $E_c$. When $\mu $ crosses below $E_c$, a crossover from BCS to BEC occurs.} 
 \label{fig:3}
\end{figure}
\begin{figure}[h]
   \includegraphics[width=8cm,angle=0]{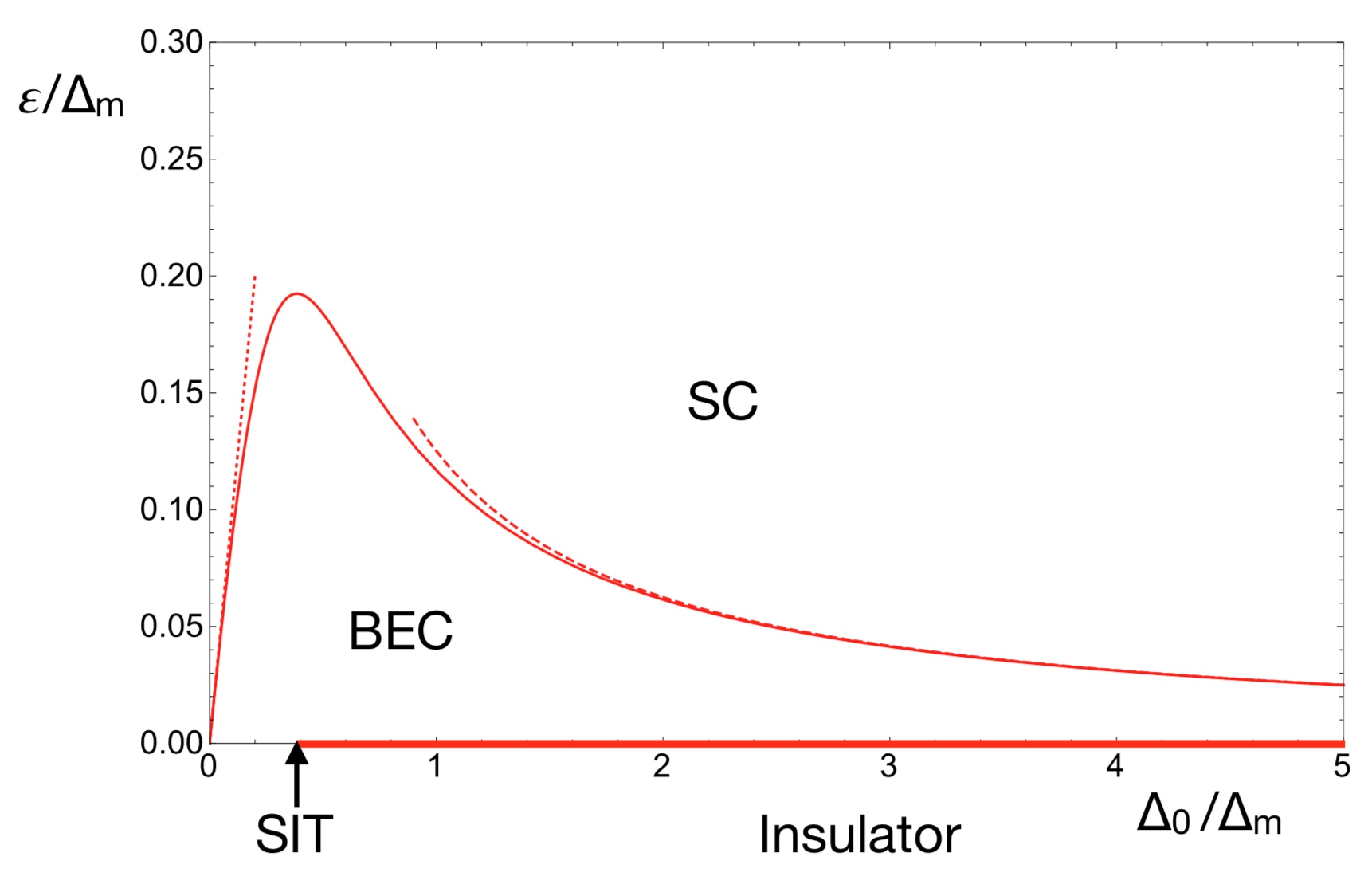}
\caption[]{ BCS-BEC crossover diagram: doping parameter $\varepsilon$ versus semiconductor band gap $\Delta_0$ in units of $\Delta_m$, as obtained by the crossover condition $\mu=\Delta_0$ for 2D DOS.  }
 \label{fig:4} 
\end{figure}
By numerically solving these equations  we obtain plots  for the superconducting order parameter $\Delta$, Fig.\ref{fig:3} (top) and the chemical potential $\mu$ Fig.\ref{fig:3} (bottom) as functions of the semiconductor gap $\Delta_{0}$. We
thereby reproduce the results of Ref. \cite{Nozieres}: in the undoped semiconductor there is 
 a sharp superconductor-insulator transition at a critical $\Delta_{0c},$
 which occurs at half of the superconducting  order parameter in a metallic superconductor, $
\Delta_{0c}= \Delta_m/2$. Here $\Delta_m$ parametrizes the strength of the attractive interaction via Eq. (\ref{ConstDOSSelfConsistencySimplified}).
  We note that this result holds in the limit of $\omega_D \gg \Delta_0$ 
  only, where the energy range of attraction extends beyond the Energy gap $\Delta_0$. In the opposite limit, the undoped  system would remain in the insulator phase. 
At finite doping, the pairing amplitude
$\Delta$ is finite for any value of the semiconducting gap $\Delta_0$, since there are always charge carriers present, which can be paired for any value of $\Delta_0$.

  As mentioned above, there is a crossover to BEC
   at low concentration of charge carriers. This 
   can be seen by the fact that as the paring sets in, 
  the chemical potential $\mu$ drops below the conduction band edge even when it has been in the conduction band before, see Fig.\ref{fig:3}(bottom).
    We obtain the correlation length  for the 2D density of states for $ D \gg \Delta_0$,
    \begin{equation} \label{xi2d}
       \xi^2 =  \frac{1}{4 m\Delta}  h(s=\frac{\mu}{\Delta},t=\frac{\Delta_0}{\Delta}),
    \end{equation}
    where 
    \begin{eqnarray}
    && h(s,t) = (\pi-\arctan (t-s) -\arctan (t+s))^{-1}  
    \langle 2 - \pi t +
    \nonumber \\ && \sum_{\alpha = \pm 1}((t-\alpha s) \arctan (t-\alpha s) 
    +\frac{1}{(t-\alpha s)^2+1}) \rangle.
    \end{eqnarray}
     For $\mu > \Delta_0$ we recover the BCS coherence length $\xi$
     given by $\xi^2 =  \frac{\mu-\Delta_0}{4 m\Delta^2} $\cite{pistolesi2}.
     When the chemical potential is at the band edge $ \mu = \Delta_0$,
     we find $\xi^2 = 1/(\pi m \Delta)$,  which is  the size of a single bound electron pair with pairing energy $\Delta$.
      For the  undoped semiconductor with symmetric bands, $\mu=0,$ 
       the coherence length is given by  $\xi^2 = 1/(3 m\Delta_0)$ for 
       $\Delta \rightarrow 0$, 
       which coincides with the size of a single bound electron pair with binding energy $\Delta_0$. 
        For $|\mu| < \Delta_0$ and $\Delta \rightarrow 0$
        one finds  $\xi^2 = 1/(3 m\Delta_0) (\Delta_0^2 +\mu^2)/(\Delta_0^2 -\mu^2).$ 
        We note that while this defines the smallest size of the bound pair in this simple two-band model with band gap $ 2 \Delta_0$, the actual size of the bound electron pair is modified by the fact that the states in the tails of the band of a doped semiconductor are localized with a finite localization length $L_c$, which in the dilute dopant limit becomes the effective Bohr radius of the ground state of the dopant levels. Thus, as the doping is reduced there occurs a metal-insulator transition to Anderson localized states, which has to be implemented in the pairing theory to obtain  a more realistic description of the BCS-BES crossover and may  result in  a localization transition to localized bosons \cite{AndersonSIT}.

    We  conclude 
    that there is a crossover from superconductivity to dilute  bound electron pairs 
     when the chemical potential is at one of  the band edges, $ \mu = \pm \Delta_0$.
     Inserting that condition into Eqs. (\ref{ConstDOSMainEqn1}),
     (\ref{ParticleConservationDopedSCSimplified1})  we find 
     $\varepsilon = \Delta_0 +\Delta/2 -\sqrt{\Delta_0^2 +\Delta^2/4},$
     where $\Delta$ is the positive solution of the quartic equation 
      $\Delta^4/\Delta_m^4 + 4 \Delta \Delta_0/\Delta_m^2 -1 =0.$
    Thereby,      the quantum phase diagram 
     in the parameter space of doping 
      $\varepsilon$ versus $ \Delta_0$, see Fig.\ref{fig:4} showing a parameter regime where Bose-Einstein condensation occurs below a critical temperature $T_c$. 
      This diagram has  already been obtained for the two-band model with a 2D density of states in Ref. \cite{Nozieres}. 
      For $\Delta_0 \gg \Delta$, one obtains that the BCS-BEC crossover occurs for  $\varepsilon/\Delta_m = 1/(8 \Delta_0/\Delta_m)$
      (dashed line in  Fig. \ref{fig:4}). 
       For $\Delta_0 \ll \Delta$, one obtains that the BCS-BEC crossover occurs for  $\varepsilon/\Delta_m = \Delta_0/\Delta_m$
      (dotted line in  Fig.\ref{fig:4}). 
   
\begin{figure}[t!] 
\includegraphics[width=7cm]{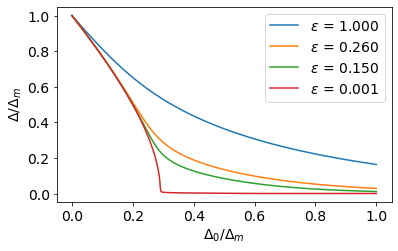}
\includegraphics[width=7cm]{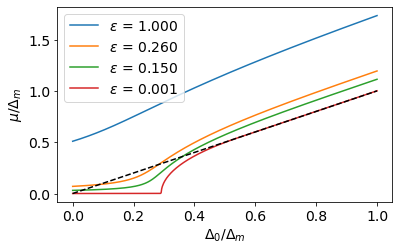}
\caption{ 
Top: superconducting order parameter $\Delta$. Bottom: chemical potential $\mu$ as function of semiconductor band gap $\Delta_{0}$, at different doping levels $\varepsilon$
 for $T=0\thinspace K$ for  3D density of states.  $\Delta_m$ is the superconducting order parameter for the metallic ($\Delta_0=0$) case,  Eq. (5).  The dashed line is the position of the conduction band edge $E_c$. When $\mu $ crosses below $E_c$, a crossover from BCS to BEC occurs.} 
 \label{fig:5}
\end{figure}
\begin{figure}[h]
   \includegraphics[width=8cm, angle=0,]{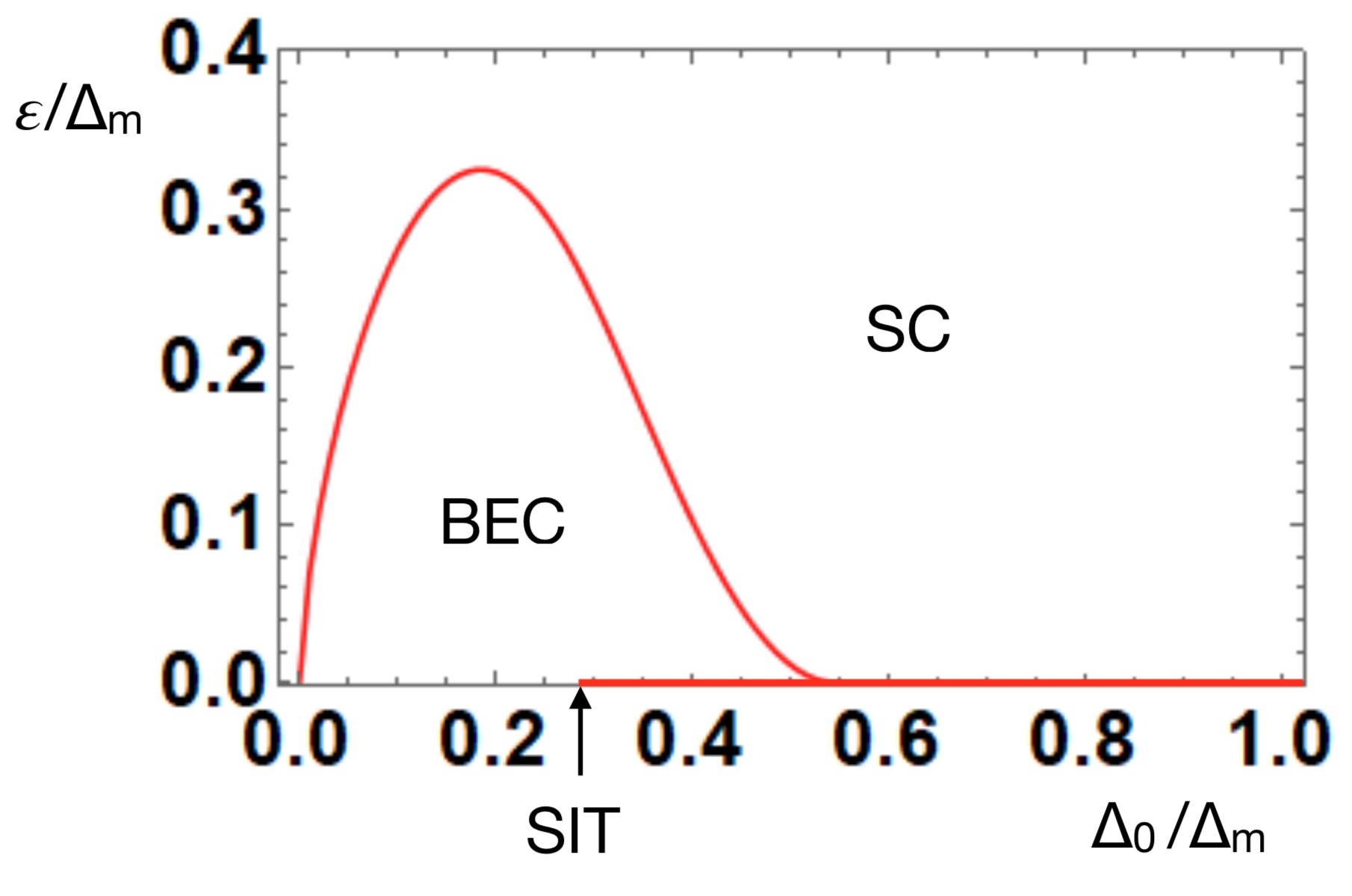}
\caption[]{ BCS-BEC Crossover diagram: doping parameter $\varepsilon$ versus semiconductor band gap $\Delta_0$ in units of $\Delta_m$ as obtained by the crossover condition $\mu=\Delta_0$  for 3D DOS. }
 \label{fig:6} 
\end{figure}

\subsection{Two-band model with 3D DOS} 

Next, we consider a density of states (DOS) which is more realistic for 
 3-dimensional semiconductors,  shown in Fig. \ref{fig:1}(right) (green):
the DOS  has a square-root dependence on the energy,
 in the conduction band,   $\rho\left(\xi_k\right)=\rho_{0c}\sqrt{\xi_k-E_{c}}$, for $E_c < \xi_k < D,$
  whereas in the valence band $\rho\left(\xi_k\right)=\rho_{0v}\sqrt{-\xi_k + E_{v}}$, for $-D < \xi_k < E_v,$
  and $\rho\left(\xi_k\right)=0$ in the band gap for $E_v < \xi_k < E_c.$
Here,  $\rho_{0c/v}=\sqrt{2m_{c/v}^{3}}/(\hbar^{3}\pi^{2}),$ 
where $m_{c/v}$ is the effective mass in the conduction/valence band, respectively.  We assume $m_c=m_v$ in the following. 
 As outlined in the Appendix the BCS equation yields then, 
  assuming that $\omega_D$ is a large energy scale, 
 Eq. \ref{EllipticEqnBCSGapEqn} and the particle conservation yields 
Eqs. \ref{EllipticEqnParticleConservation}.
This  defines the  set of equations that model the three-dimensional BCS superconducting semiconductors yielding the order parameter $\Delta$
and the chemical potential $\mu$.
We solve these equations numerically to obtain  the  superconducting order parameter $\Delta$, Fig. \ref{fig:5}(top), and chemical potential $\mu$, Fig. \ref{fig:5}(bottom) as functions of the semiconductor gap $\Delta_{0}$, the attractive interaction $U$ via $\Delta_{m}=2\omega_{D}\exp\left(-1/(\rho U)\right)$, and the doping parameter $\varepsilon/\Delta_m$. 
Without doping, $\varepsilon =0$, the superconducting order parameter $\Delta$ drops to zero when the semiconductor gap reaches the critical value $ \Delta _{0c} = 0.29 \Delta_m$.
 Thus, this superconductor-insulator transition occurs already at a smaller semiconducting band gap,  than for the step function DOS, as expected, since the density of states is smaller when approaching the band edges compared to the 2D case, and thus less quasiparticles are available to pair and participate in the condensate. 
   For finite $\omega_D$, there  would only be a superconducting phase
    when  $\omega_D > \Delta_0.$ 
 For finite doping $\varepsilon$, the order parameter $\Delta$ persists 
 for all values of the semiconductor gap $\Delta_0$, but is  for the same values of 
  $(\Delta_0,\varepsilon)$ substantially smaller than for the step function DOS. 
 As discussed in the previous section, the
condition $\mu= \pm \Delta_0$ gives  the BCS-BEC crossover line in  parameter space spanned by
the 
doping parameter $\varepsilon$ and the semiconductor gap  $\Delta_0$. In Fig. \ref{fig:6} we plot the resulting phase diagram as obtained by a numerical solution of the above equations  for the 3D DOS for $\mu =\Delta_0$.  Remarkably, we find that 
for large  semiconductor band gaps $\Delta_0 >  \Delta_0^* = 0.5 \Delta_m$, there is no solution with  $\mu =\Delta_0$ for
finite doping $\varepsilon > 0,$  within the numerical accuracy of at least $10^{-4}$, so that there exists  no BEC, but rather a direct transition to a BCS superconductivity phase  as shown in Fig. \ref{fig:6}.

        \section{Spatial variation along a superconducting p-n junction}
      
\begin{figure}[t]
\includegraphics[width=0.5\textwidth,angle=0]{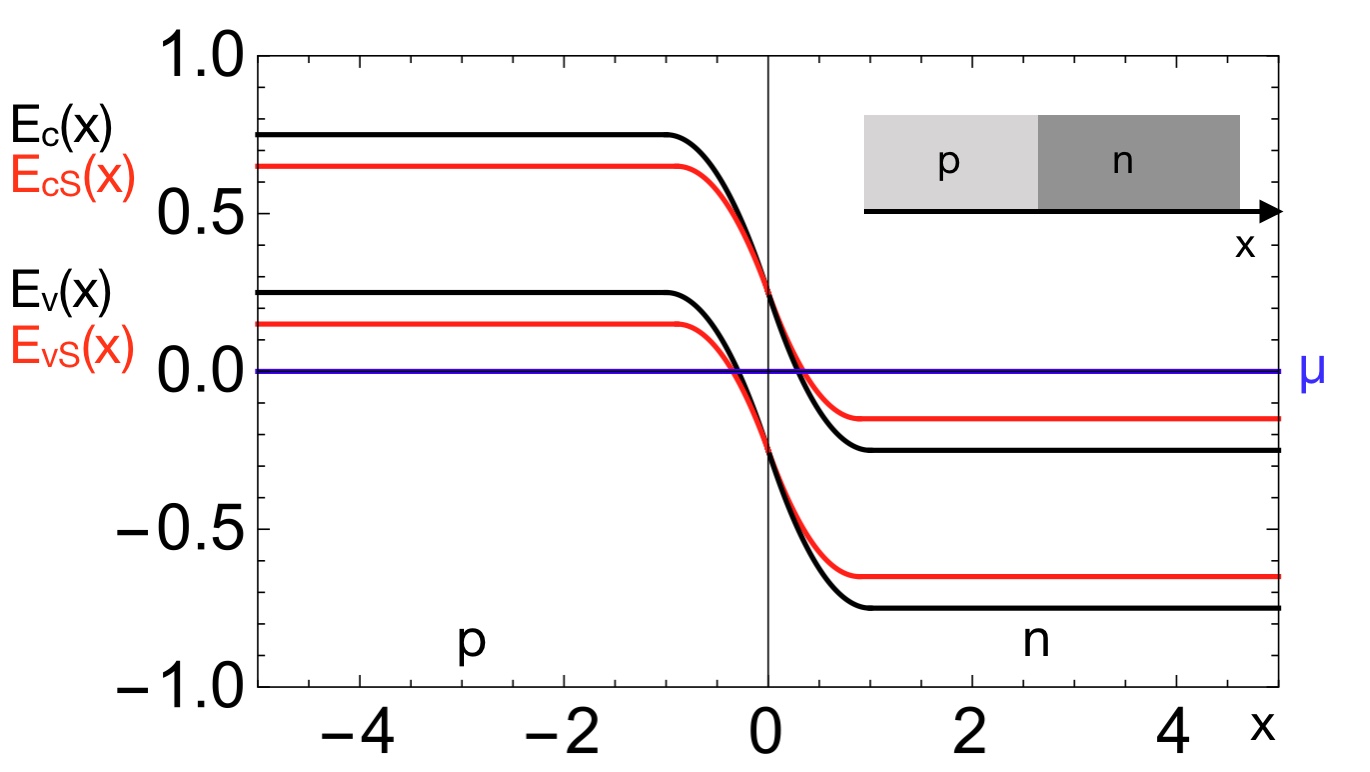}
\caption{ Energy Band diagram of p-n junction with spatial variation of band edges $E_C(x),E_V(x)$ (black).  Superconductivity caused by the attractive interaction shifts 
the band edges  to  $E_{CS}(x),E_{VS}(x)$ (red). The chemical potential remains constant without external bias (blue). Inset: geometry of the p-n junction.  }
\label{fig:7}
\end{figure}

      Having derived the superconducting order parameter $\Delta$ and  chemical potential $\mu$   as functions of the semiconductor gap $\Delta_0$  and the doping level
      $\varepsilon$, we can 
       study the  effect of pairing on the properties of p-n junctions  in the presence of an attractive interaction $U$.
       For  doping levels $\varepsilon_{n,p}$ 
the potential drop    across a conventional   p-n junction is  given by 
\begin{align}\label{eDelPhi}
e\,\Delta\phi =\varepsilon_{n}+\varepsilon_{p}+2\Delta_{0}.
\end{align}
The charge density drops in the  depletion region which has on  the n-side a width
$d_{n}$ and  on the p-side the depletion width $d_{p}$. Using Poisson's
equation,
$
\frac{d^{2}\phi}{dx^{2}}=\varrho(x)/\epsilon,
$
where $\varrho\left(x\right)$ is the charge density and $\epsilon$
is the dielectric constant, one finds in the depletion approximation, which assumes, when solving the Poisson equation,  that there are no charge carriers in the depletion region, 
\begin{align}\label{-ePhi}
-e\phi\left(x\right)= \frac{e\Delta\phi}{N_D+N_A}\cdot
\begin{cases}
-N_{A}, x> d_n,\\
-N_{A}(1-(\frac{x}{d_{n}}-1)^{2}), d_{n}>x >0,\\
N_{D}(1-(\frac{x}{d_{p}}+1)^{2}),-d_{p}<x<0,\\
N_{D}, x < - d_{p}.
\end{cases}
\end{align}
Here, the depletion lengths in the $n,p$ regions are respectively given by
 $d_{n/p} = (N_{A/D}/N_{D/A} \epsilon \Delta \phi/(2 \pi e (N_D+N_A)))^{1/2},$ where $\epsilon$ is the bulk dielectric constant 
  of the semiconductor. 
For a given energy gap of the semiconductor $\Delta_0$, we thus obtain  the spatial variation of the conduction and valence band edges across the p-n junction,
\begin{align}\label{CB}
E_{C}(x)=-e\phi\left(x\right)+\Delta_{0},
E_{V}(x)=-e\phi\left(x\right)-\Delta_{0},
\end{align}
as plotted in Fig. \ref{fig:7} (black lines). 
The electrochemical potential is given by $\mu_{em}(x) = \mu +e \phi(x).$
For simplicity we assume that both, the n- and p-sides are equally doped,
$\varepsilon_{n}=\varepsilon_{p} = \varepsilon$, $N_A=N_D$, $d= d_n=d_p=d
=  (\epsilon \Delta \phi/(4 \pi e N))^{1/2}$.
As we consider the p-n junction without an external bias, the chemical potential 
$\mu$ remains independent of the position $x$ across the junction, $\mu=0$ (blue line in  Fig. \ref{fig:7}). 

Turning  on superconductivity takes charge carriers into the condensate, 
 changing the electrochemical energy on  both sides of the junction 
 by an amount which equals  the
 superconducting binding energy. 
Thereby, 
the potential energy drops  across the p-n junction  in the presence of superconductivity  by an amount given by 
\begin{align}\label{eDelPhi}
e\,\Delta\phi_S = e\phi_S(x  \ll -d_p)  - e\phi_S(x \gg d_n)
\nonumber \\ = 2 \mu_{em} (\Delta_0, \varepsilon, \Delta_m),
\end{align}
where  the parameters $\Delta_0, \varepsilon, \Delta_m$ 
 are the semiconductor band gap, the doping level and the 
  superconducting order parameter in the metallic limit, as defined above.

This change of the potential drop changes 
 the spatial dependence of the  potential 
 $\phi_S(x)$, accordingly,  resulting in the new spatial variation of the band edges, 
\begin{align}\label{CB_new}
E_{C{S}} (x)=-e\phi_{S}\left(x\right)+\Delta_{0},
E_{V{S}} (x)=-e\phi_{S}\left(x\right)-\Delta_{0}.
\end{align}
In depletion approximation this yields
\begin{align}\label{-ePhis}
-e\phi\left(x\right)= \frac{e\Delta\phi^S}{2}\cdot
\begin{cases}
-1, x> d^s,\\
-1 +(\frac{x}{d^s}-1)^{2}, d^s>x >0,\\
1-(\frac{x}{d^s}+1)^{2},-d^s<x<0,\\
1, x < - d^s,
\end{cases}
\end{align}
with the depletion width reduced to  $d^s
=  (\epsilon \Delta \phi^s/(4 \pi e N))^{1/2}$.

 The spatial variation of 
 $\Delta(x)$ at junctions can be derived from the Gorkov equations\cite{Larkin1966}, or equivalently from the Bogoliubov-de Gennes equations\cite{Degennes1964},\cite{Spuntarelli2010}. For Josephson contacts, 
 such as junctions of  superconductors 
 with an insulating oxide layer in between, it was found that 
   $\Delta(x)$ varies  in close vicinity of the junction
    on length scales of the order of the insulator thickness,
    as imposed by the drop of the charge density in the oxide layer.
    Further away from the junction,  however, 
      $\Delta(x)$ varies on length scales of the order of the bulk coherence length 
      $\xi$\cite{Degennes1964},\cite{Spuntarelli2010}, since
      the variations on shorter length scales in in the bulk superconductor energetically suppressed by 
       long range order.
      Thus, when the coherence length is larger than the 
       depletion length, $\xi = v_F/\Delta > d$, we can assume that the spatial variation of 
 $\Delta(x)$ at the p-n junction is dictated by the electrostatics at the junction, and thereby the reduced charge carrier density as parameterised by the
 the electrochemical potential $\mu_{em} (x)$. 
 While the  chemical potential  $\mu$
is constant in the p-n junction without external bias, the chemical potential entering in
the 
pairing  equation
  Eq. (\ref{ConstDOSMainEqn1})  for the 2D system,  and in Eq. 
 (\ref{EllipticEqnBCSGapEqn}) for the 3D system 
 is rather the electrochemical potential
$\mu_{em} (x)$ as measured relative to the middle of the semiconductor gap at the respective position x,
 which is for  $\mu=0$ in depletion approximation given  by 
 \begin{align}\label{mu_ems}
\mu_{em}^s(x) = \mu_{em} (\Delta_0, \varepsilon, \Delta_m)) \cdot
\begin{cases}
1, x> d^s,\\
1 -(\frac{x}{d^s}-1)^{2}, d^s>x >0,\\
-1+(\frac{x}{d^s}+1)^{2},-d^s<x<0,\\
-1, x < - d^s,
\end{cases}
\end{align}
  Therefore, to get the spatial variation of 
 $\Delta(x)$ on length scale $d$
 along the length of the p-n junction  for different values of $\Delta_{0}$ and $\varepsilon$,  we can 
  in a first,  local density approximation, insert  $\mu_{em} (x)$ as given by 
 Eq. (\ref{mu_ems}) 
 into the  pairing  equation
  Eq. (\ref{ConstDOSMainEqn1})  for the 2D system,  and in Eq. 
 (\ref{EllipticEqnBCSGapEqn}) for the 3D system   and solve for  $\Delta(x)$
for every position $x$.
 
{\it  p-n junction of 2D systems.} For the 2D system we find thereby  two different kinds of superconducting p-n junctions when the bulk is in the BCS phase: 
\begin{figure}[t]
\includegraphics[width=0.5\textwidth,angle=0]{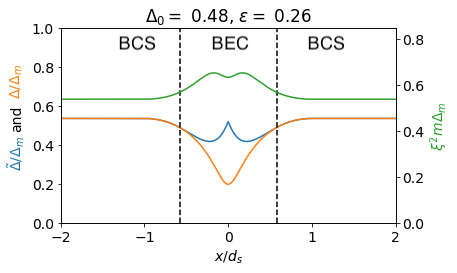}
\includegraphics[width=0.5\textwidth,angle=0]{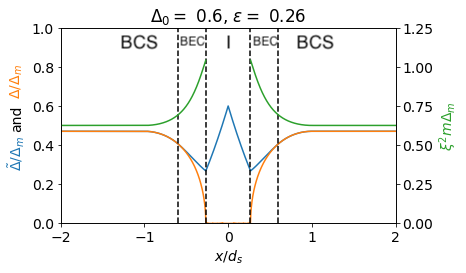}
\caption{Order parameter $\Delta\left(x\right)$, quasiparticle gap $\Tilde{\Delta}(x)$, and the coherence length $\xi(x)$ across two types of 2D p-n junctions. (Top) BCS-BEC-BCS with semiconductor gap $\Delta_{0}=0.48 \Delta_{m}$ and doping 
 $\varepsilon = 0.26  \Delta_m$. (Bottom) BCS-BEC-I-BEC-BCS with semiconductor gap $\Delta_{0}=0.6 \Delta_{m}$ and doping 
 $\varepsilon = 0.26  \Delta_m$.}
\label{fig:8}
\end{figure}               

 
  1. {\bf BCS-BEC-BCS junction}: For $\Delta_0 < \Delta_m/2$
  the order parameter $\Delta\left(x\right)$ decreases in the space charge region, but remains finite with a minimum in the middle of the pn-junction, as shown in Fig. \ref{fig:8} (Top).
  However, we find that even when the bulk system is in the BCS superconducting phase, there emerges a BEC layer at the pn-junction
 as the chemical potential 
  moves into the band gap at the pn-junction. This BEC condensate  extends throughout the pn-junction in a regime  of width $d_{BEC} = 2 d_s (1- \sqrt{1-\Delta_0/\mu_{\rm em}}),$
  as obtained by the condition $\mu_{\rm em} (x = \pm d_{BEC}/2)  = \pm \Delta_0.$ The quasiparticle  excitation gap $\Tilde{\Delta}(x)$ remains 
  for the condition $\Delta_0 < \Delta_m/2$ finite throughout the pn-junction, decreasing first as the order parameter  $\Delta\left(x\right)$ decreases, reaching a minimum and increasing again, as the 
  chemical potential moves in the middle of the semiconductor band gap. 
    Interestingly, the coherence length, 
    which we calculate approximately using Eq. (\ref{xi2d}), 
   in the BCS phase 
     increases with the decrease of $\Delta\left(x\right)$, but converges
      to a finite value in the BEC phase, and decreases to a minimum, 
       in the middle of the pn-junction.

  2. {\bf BCS-BEC-I-BEC-BCS junction}: For $\Delta_0 > \Delta_m/2$
  the order parameter $\Delta\left(x\right)$ is found to decrease in the space charge region to $0$ , as shown in Fig. \ref{fig:8} (Bottom), with a finite layer of an insulator phase in the middle of the junction. 
  Thus, as the chemical potential 
  moves into the band gap at the pn-junction, there  is   a BEC condensate at each of  the two  surfaces of the p-n junction, each  of finite width $d_{BEC} =  d_s ((1- \frac{\Delta_0}{\mu_{\rm em}} \sqrt{1- \Delta_m^2/(4 \Delta_0^2)} )^{1/2}- (1-\frac{\Delta_0}{\mu_{\rm em}})^{1/2}),$ separated by an insulating layer, where $\Delta=0$. The quasiparticle  excitation gap $\Tilde{\Delta}(x)$ remains finite throughout the pn-junction, decreasing first as the order parameter  $\Delta\left(x\right)$ decreases, reaching a minimum at the boundary between the BEC and the insulator phase and increasing again in the insulator layer, as the  
  chemical potential moves in the middle of the semiconductor band gap. 
    Interestingly, the coherence length, 
    as approximated  with Eq. (\ref{xi2d})
    which in the BCS phase 
     increases with the decrease of $\Delta\left(x\right)$, converges
      to a finite value at the boundary between the  BEC and the insulator phase, where the order parameter vanishes. 
       
  \begin{figure}[t]
\includegraphics[width=0.43\textwidth,angle=0]{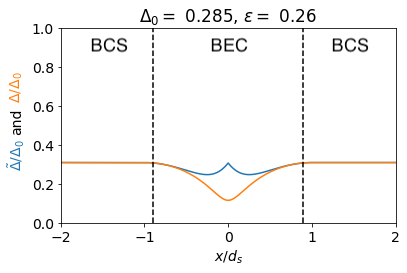}
\includegraphics[width=0.43\textwidth,angle=0]{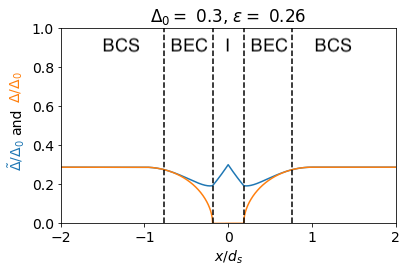}
\includegraphics[width=0.43\textwidth,angle=0]{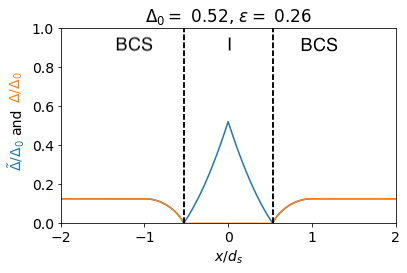}
\caption{The spatial variation of the order parameter $\Delta\left(x\right)$ and the  quasiparticle gap $\Tilde{\Delta}(x)$ across  the 3D  p-n junction. (Top) BCS-BEC-BCS junction with semiconductor gap $\Delta_{0}=0.285\Delta_{m} <\Delta_{0c}$ and doping $\varepsilon = 0.26 \Delta_m$. (Center) BCS-BEC-I-BEC-BCS junction  with semiconductor gap $\Delta_{0}=0.3\Delta_{m} > \Delta_{0c}$ and doping $\varepsilon = 0.26 \Delta_m$. (Bottom) BCS-I-BCS junction with semiconductor gap $\Delta_{0}=0.52\Delta_{m}$ and doping $\varepsilon = 0.26 \Delta_m$, showing 
the appearance of gapless quasiparticle excitations at the boundary between the BCS and the insulator phase. }
\label{fig:9}
\end{figure}
  
  {\it  p-n junction of 3D systems.}
  In the 3D systems  we   find, when the bulk is in the BCS phase,  a BEC layer at the pn-junction occurs only for sufficiently small semiconductor gaps
   $\Delta_0 <\Delta_0^* $.
  Thus, we 
  find in 3D  three  different kinds of superconducting p-n junctions,
   when the bulk is in the BCS phase: 
  
  1. {\bf BCS-BEC-BCS junction}: 
  For small semiconductor gaps $\Delta_0 < \Delta _{0c} =   = 0.29 \Delta_m$
  the order parameter $\Delta\left(x\right)$ decreases in the space charge region, but remains finite with a minimum in the middle of the pn-junction, as shown in Fig. \ref{fig:9} (Top). Thus, as the chemical potential 
  moves into the band gap at the pn-junction, there  appears  a BEC condensate, 
   where the chemical potential is outside of  the band edges, which extends throughout the pn-junction in a regime  of width $d_{BEC},$
  as obtained by the condition
  $\mu_{\rm em} (x = \pm d_{BEC}/2)  = \pm \Delta_0.$ The quasiparticle  excitation gap $\Tilde{\Delta}(x)$ remains finite throughout the pn-junction, decreasing first as the order parameter  $\Delta\left(x\right)$ decreases, reaching a minimum and increasing again, as the 
  chemical potential moves into the middle of the semiconductor band gap.

  2. {\bf BCS-BEC-I-BEC-BCS junction}:  For large semiconductor band gaps $\Delta_0 > \Delta _{0c} =   = 0.29 \Delta_m$
  the order parameter $\Delta\left(x\right)$ decreases in the space charge region to $0$ , as shown in Fig. \ref{fig:9} (Center), with a finite layer of an insulator phase in the middle of the junction. 
  Thus, as the chemical potential 
  moves into the band gap at the pn-junction, there  is   a BEC condensate at each of  the two  surfaces of the p-n junction, each  of finite width $d_{BEC},$ separated by an insulating layer, where $\Delta=0$. The quasiparticle  excitation gap $\Tilde{\Delta}(x)$ remains finite throughout the pn-junction, decreasing first as the order parameter  $\Delta\left(x\right)$ decreases, reaching a minimum at the boundary between the BEC and the insulator phase and increasing again in the insulator layer, as the 
  chemical potential moves into the middle of the semiconductor band gap. 
   
  3.  {\bf BCS-I-BCS junction}: 
 For still larger semiconductor band gaps,
 $\Delta_0 > \Delta _{0}^* =   = 0.5 \Delta_m$, there is no BEC layer anymore, 
  the order parameter $\Delta\left(x\right)$ decreases to zero as the chemical potential reaches the band edge, as shown in Fig. \ref{fig:9} (Bottom), 
  reaching directly  an insulator phase as the chemical potential 
  moves into the band gap at the pn-junction. Remarkably, the quasiparticle  excitation gap $\Tilde{\Delta}(x)$  vanishes at the boundary of the space charge region, decreasing first to zero as the order parameter  $\Delta\left(x\right)$ decreases to zero,  and increasing again in the insulator layer, as the 
  chemical potential moves into the middle of the semiconductor band gap. Thus, there appear gapless 
   quasiparticle excitations at the boundary to the space charge region. 
   
\section{Conclusions and Discussion} 
   
   Thus, we have shown that in superconducting pn-junctions there can  appear layers of BEC condensates even when the bulk is in the BCS state.
   This opens the possibility to create layers of  BEC   and study their properties  in the well controlled setting of doped semiconductors, where the doping level can be varied to  change and control the thickness of BEC and insulator layers. 
    The BEC condensate can be detected by scanning tunneling microscopy, 
     where instead of the sharp  coherence peaks in the BCS phase, a maximum in the tunneling density of states in the band  which is closest to the chemical potential, is expected, as plotted in Fig. \ref{fig:3} (Bottom). Also, 
      the fact that the quasiparticle excitation gap remains finite throughout the pn-junction when there is a BEC layer, while there are gapless excitations in a conventional BCS-I-BCS junction, might be amenable to experimental detection.

      Moreover, attaching sufficiently small leads in lateral direction, the superconducting pn-junction may enable one to study the transport properties of the BEC layers directly. 
      
      As qualitatively outlined in Ref. \onlinecite{Mannhart}, the 
      superconductor critical current $I_c$ is 
       expected to be still  dominated by the bulk superconducting order parameter 
        $\Delta$ and the normal small voltage resistance of the pn-junction $R_n$, as in a conventional Josephson contact, 
         yielding for identical $\Delta$ on both sides of the junction 
        $I_c R_n = \pi \Delta/(2e).$ The presence of a BEC layer might modify that 
         product due to the spatial variation of the  order parameter, and the quasiparticle exciation gap, 
          see Figs. \ref{fig:8},\ref{fig:9}.  We will leave the derivation  as a task for further studies. 
   
 For a conventional semiconductor with $ \epsilon = 10$, $ \Delta \Phi = 1V$  and $N_D=N_A= 10^{18} 
cm^{ -3}$, the depletion width is 
$ d \approx  50 nm$ \cite{Mannhart}, whereas 
 in p-n junctions of   cuprate semiconductors 
 $ \Delta \Phi$ can be several volts, $N_D=N_A= 5. \times 10^{21} 
cm^{ -3}$, yielding only  $ d \approx  1 nm$ which is the same order as the thickness of oxide barriers in typical Josephson junctions.  Indeed, cuprate semiconductors with 
 a superconducting phase for both hole and electron doping have been found, see  Ref. \onlinecite{norman} for a review, which may therefore be realisations of  homogeneous p-n junctions, 
where we can expect BEC layers of the thickness of the order of $d_{BEC} \approx 1nm$.

 The theory can be extended  to hetero-junctions with two different host materials with different band gaps on the n- and p- doped side of the junctions, resulting in band discontinuities at the junction to study  what effect this   has on the existence of a BEC layer.

The  $I(V)$ characteristics of superconducting 
pn junctions has been 
discussed qualitatively  in Ref. \onlinecite{Mannhart}.
 We leave it for future work to extend our theory to include a potential difference and thereby allow a quantitative derivation of current voltage characteristics, and 
    to study what consequence BEC-layers have for the I(V)-characteristics.

Recently, Josephson junctions in the BCS-BEC crossover range have 
been reviewed in Ref. \onlinecite{Spuntarelli2010}  by solving the 
Bogoliubov-de-Gennes equations for this problem. These authors  did not discuss the appearance of a BEC
layer  at the junction when the bulk is in the BCS phase. 
However, we expect, that, since the carrier concentration is reduced  in the vicinity of an oxide layer,  a BEC layer may  also appear at such  BCS- Josephson junctions with an oxide layer, a question  we leave for future research. 

 An extension of the Bogoliubov-de-Gennes equations\cite{Degennes1964},
 \cite{Spuntarelli2010} to the 2-band model and its application 
 to the superconducting p-n junctions will lead also further insights 
  into the spatial variation of the order parameter, when solved
   self consistently with the Poisson equation. This calculation, where the condensation amplitude as well as the charge redistribution are self-consistently computed can be performed within the tight binding framework \cite{ghosal2001, black-shaffer2008, rai2019}. In particular, one can expect 
   deviations from our result for the spatial change of
    $\Delta(x)$ on length scales  of the order of the bulk coherence length 
     $\xi$. Also, additional discrete states might appear 
     as solutions of the  Bogoliubov-de-Gennes equations
     at the junction, similar to the Andreev bound states  found in 
     Josephson junctions \cite{Spuntarelli2010}. This raises interesting questions  for future research, as the change from electron like to hole like charge carriers across the junction challenges the conventional interpretation of Andreev bound states. 


In  our study  we have assumed  zero temperature $T=0K$, and it remains to be  extended  to finite temperatures $T$. 
Furthermore,  while our study  employs the   mean field approximation of the many body physics,
the effect of fluctuations of the order parameter amplitude and phase  need to be  included  to  get a better 
  understanding of the stability of 
 the long range order at finite temperature and in the thin film,  2D limit\cite{Kosterlitz1973},\cite{Nozieres},
\cite{Larkin2005}. 
  
The disorder introduced by the dopants will furthermore lead to Anderson localization of charge carriers and accordingly may result in a layer of disorder localized Bosons at the p-n junction, 
reducing the thickness of the extended BEC layer. 
These issues will be  subject  for future research.


\acknowledgments

S.K. gratefully acknowledges support from DFG KE-807/22-1. This work was supported by the US Department of Energy under grant number DE-FG03-01ER45908. The numerical computations were carried out on
the University of Southern California High Performance
Supercomputer Cluster.

 \appendix
 
\section{  3D 2-band model}
 Following an  approach similar to Eagles \cite{Eagles}, who  solved the  BCS equation and particle conservation  equation
  for a single-band semiconductor, 
  we rearrange the particle conservation equation Eq. \ref{ParticleNumberConservation} of the two-band model to get
\begin{equation}\label{ParticleConservation2BandModel}
\frac{2}{3}\thinspace\left(\frac{\varepsilon}{\Delta}\right)^{\frac{3}{2}}=Q\left(\lambda_{1}\right)-Q\left(\lambda_{2}\right),
\end{equation}
where
$
Q\left(\lambda_{i}\right)={\int_{0}^{\infty}x^{2} dx (1-(x^{2}-\lambda_{i})/\sqrt{1+\left(x^{2}-\lambda_{i}\right)^{2}})}
$for $i=1,2.$
with
$\lambda_{1}\equiv (\mu-\Delta_{0})/\Delta$, and $\lambda_{2}\equiv (-\Delta_{0}-\mu)/\Delta$. Here, we changed the integration  parameters
 to $x^2= (\xi - \Delta_0)/\Delta$ for 
  $i=1$ and  $x^2= (\xi + \Delta_0)/\Delta$ for 
  $i=2$. We  approximated  $D/\Delta \rightarrow \infty.$
 As  in the 2D limit,  we  assume   that $\omega_D$ is a  large energy scale, for simplicity. This means that we 
 assume that the  electron-electron  interaction is attractive in both bands.
 Therefore, we  can set $\omega_D = D$ so that 
the BCS self-consistency equation  simplifies to 
\begin{equation}\label{BCSGapEqn2BandModel}
1\simeq \rho_{0}U\Delta^{\frac{1}{2}}\left[P\left(\lambda_{1}\right)+P\left(\lambda_{2}\right)+\sqrt{\frac{4\omega}{\Delta}}\right],
\end{equation}
where
$P(\lambda_{i})={\int_{0}^{\infty}dx (x^{2}/\sqrt{1+\left(x^{2}-\lambda_{i}\right)^{2}}-1)},$ $i=1,2.$
We follow the approach by Pistolesi \cite{Pistolesi} to rewrite Eqs. \ref{ParticleConservation2BandModel} and \ref{BCSGapEqn2BandModel} in terms of elliptical integrals and obtain for the equation ensuring particle conservation
\begin{align}\label{EllipticEqnParticleConservation}
\frac{2}{3}\thinspace\left(\frac{\varepsilon}{\Delta}\right)^{\frac{3}{2}}= 
\sum_{i=1,2} \sigma_i
\lambda_{i}\left(1+\lambda_{i}^{2}\right)^{\frac{1}{4}}E\left(\frac{\pi}{2},k_{i}\right)+ \notag
 \\
\sum_{i=1,2} \sigma_i \frac{  (1+\lambda_i^2 )^{1/4} }{ 2 ( \lambda_{i}+\sqrt{1+\lambda_{i}^{2} })} F (\frac{\pi}{2} , k_i ),
\end{align}
where $\sigma_1=1$, $\sigma_2=-1$  and
$k_{i}^{2}=\frac{\sqrt{1+\lambda_{i}^{2}}+\lambda_{i}}{2\sqrt{\left(1+\lambda_{i}^{2}\right)}}$ for $i=1,2$.
Here, $F(\varphi,k)$ are $E(\varphi,k)$ the incomplete elliptic integral of the  first and second kind, respectively. 
The pairing equation becomes 
\begin{align}\label{EllipticEqnBCSGapEqn}
1=2\rho_{0}U\sqrt{\Delta} \sum_{i=1,2}
\biggl[
-\left(1+\lambda_{i}^{2}\right)^{\frac{1}{4}}E\left(\frac{\pi}{2},k_{i}\right)+ \notag
\\
\frac{F\left(\frac{\pi}{2},k_{i}\right)}{2\left(1+\lambda_{i}^{2}\right)^{\frac{1}{4}}}
\left( \lambda_{i} + 
\frac{1}{\sqrt{1+\lambda_{i}^{2}}+\lambda_{i}} \right)+
\sqrt{\frac{\omega_D}{\Delta}}\biggr].
\end{align}
We denote  the metallic limit by   $\Delta (\Delta_0=0)=\Delta_{m},$  as defined by   Eq.  (\ref{EllipticEqnBCSGapEqn}),
when substituting there $\Delta_0=0$, $\Delta$ by $\Delta_m$ and 
$ \lambda_i$ by 
$\widetilde{\lambda}_{1} = \lambda_i|_{\Delta_0=0,\Delta=\Delta_m}  $ and
$\widetilde{k}_{i}^{2}=\frac{\sqrt{1+\widetilde{\lambda}_{i}^{2}}+\widetilde{\lambda}_{i}}{2\sqrt{1+\widetilde{\lambda}_{i}^{2}}}$ for $i=1,2$.
Since we assume that the  local attraction $U$ between the fermions does not depend on $\Delta_0$, we can equate the right hand side of Eq. \ref{EllipticEqnBCSGapEqn}  with finite $\Delta_0$ to the one obtained 
 in the metallic limit. This equation gives together with 
Eqs. \ref{EllipticEqnParticleConservation} the new set of equations that model the three-dimensional BCS superconducting semiconductors.
  


\end{document}